\documentclass[runningheads]{llncs}

\usepackage[T1]{fontenc}
\def\doi#1{\href{https://doi.org/\detokenize{#1}}{\url{https://doi.org/\detokenize{#1}}}}

\usepackage{hyperref}

\usepackage{graphicx}
\usepackage{siunitx}
\usepackage{listings}
\begin{document}

\title{Physiology-based simulation of the retinal vasculature enables annotation-free segmentation of OCT angiographs}
\titlerunning{Simulation of the retinal vasculature for annotation-free OCTA segmentation}

\author{Martin J. Menten\inst{1,2} \and
Johannes C. Paetzold\inst{1,2} \and
Alina Dima\inst{1} \and
Bjoern H. Menze\inst{3} \and
Benjamin Knier\inst{4} \and
Daniel Rueckert\inst{1,2}}

\authorrunning{M. J. Menten et al.}

\institute{Lab for AI in Medicine, Klinikum rechts der Isar, Technical University of Munich \and
BioMedIA, Department of Computing, Imperial College London \and
Department of Quantitative Biomedicine, University of Zurich \and
Department of Neurology, Klinikum rechts der Isar, Technical University of Munich}

\maketitle

\begin{abstract}

Optical coherence tomography angiography (OCTA) can non-invasively image the eye's circulatory system. In order to reliably characterize the retinal vasculature, there is a need to automatically extract quantitative metrics from these images. The calculation of such biomarkers requires a precise semantic segmentation of the blood vessels. However, deep-learning-based methods for segmentation mostly rely on supervised training with voxel-level annotations, which are costly to obtain.

In this work, we present a pipeline to synthesize large amounts of realistic OCTA images with intrinsically matching ground truth labels; thereby obviating the need for manual annotation of training data. Our proposed method is based on two novel components: 1) a physiology-based simulation that models the various retinal vascular plexuses and 2) a suite of physics-based image augmentations that emulate the OCTA image acquisition process including typical artifacts.

In extensive benchmarking experiments, we demonstrate the utility of our synthetic data by successfully training retinal vessel segmentation algorithms. Encouraged by our method's competitive quantitative and superior qualitative performance, we believe that it constitutes a versatile tool to advance the quantitative analysis of OCTA images.

\end{abstract}

\section{Introduction}

Optical coherence tomography angiography (OCTA) allows the non-invasive acquisition of high-fidelity volumetric images of the eye's circulatory system \cite{spaide2015retinal}. It has become an important tool for ophthalmologists to diagnose and monitor ocular diseases that manifest themselves as pathological changes to the eye's vessels \cite{joussen2007retinal}. Furthermore, the retina has been described as ``window to the body'' as alterations to its vasculature have been linked to various neurological and cardiac diseases \cite{liew2011retinal,london2013retina}. In order to improve the understanding and treatment of these conditions, it is crucial to automatically extract quantitative metrics that characterize the retinal vasculature. The calculation of many of these biomarkers requires a precise semantic segmentation of the blood vessels.

\begin{figure}[htbp]
\centering
\includegraphics[width=1.0\textwidth]{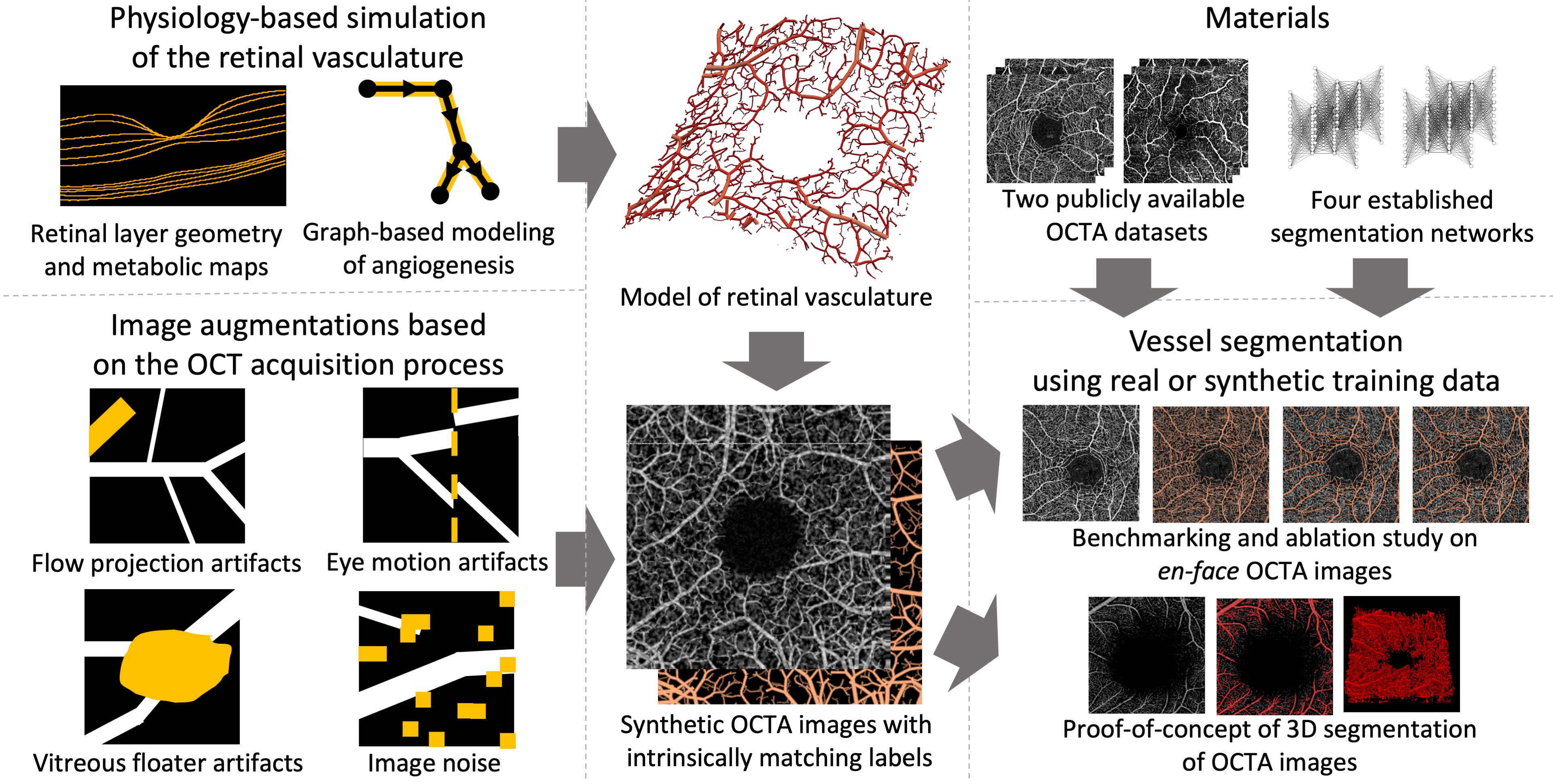}
\caption{Overview of the proposed method and conducted experiments.}
\label{fig:graphical_abstract}
\end{figure}

Recently, deep learning has seen wide-ranging success in the semantic segmentation of medical images, including the segmentation of blood vessels in angiographs. However, training of these algorithms is mostly based on large sets of annotated data, which are time-consuming to curate. Creating voxel-level annotations of the complex branching retinal vasculature in OCTA images is particularly cumbersome. Until now, only two publicly available OCTA datasets with voxel-wise annotations have been released \cite{li2020ipn,ma2020rose}. With 226 and 500 images, respectively, these datasets are several orders of magnitude smaller than the largest medical image datasets and a dataset with three-dimensional annotations has not been published yet. There is a clear need for new strategies that are independent of manual labeling to further advance the quantitative analysis of OCTA images. 

To address this need, we present a pipeline that can synthesize large amounts of realistic OCTA images with intrinsically matching ground truth labels (see figure \ref{fig:graphical_abstract}). Our proposed method is based on two novel components: 1) a physiology-based simulation that models the various retinal vascular plexuses and 2) a suite of physics-based image augmentations that emulate the OCTA image acquisition process including typical artifacts.

We evaluate the utility of our synthetically generated data for the training of supervised vessel segmentation algorithms. In extensive benchmarking experiments involving four neural network architectures and two publicly available OCTA datasets, we compare the segmentation results when training on synthetic OCTA image-label pairs versus real OCTA images with expert-derived annotations. Additionally, we quantify the intrinsic scalability of our approach and investigate how it can facilitate segmentation of the retinal vasculature in three-dimensional OCTA images.

\section{Related works}
\label{sec:relatedworks}

\noindent\textbf{Deep-learning-based segmentation of OCTA images} Similar to most segmentation tasks, convolutional neural networks constitute the top-performing methods for segmentation of tubular structures \cite{garcia2018survey}. However, the adoption of supervised deep learning for retinal vessel segmentation in OCTA images has been limited due to a lack of annotated data \cite{li2020image,li2020ipn,ma2020rose}. Beyond supervised learning, Liu \textit{et al.} have recently presented an unsupervised method for OCTA segmentation that does not rely on expert-derived labels \cite{liu2020variational}. However, their method requires multiple scans of the same subject acquired with scanners from different manufacturers, which are usually unobtainable in clinical practice.

\noindent\textbf{Learning with synthetic data} One way to overcome data sparsity in machine learning is transfer learning from synthetically generated data. Here, models are pretrained on artificial data and then refined on a small set of real image-label pairs. In medical image segmentation, this approach has been successfully applied to the segmentation of brain tumors \cite{lindner2019using}, endotracheal tubes \cite{frid2019endotracheal} and catheters \cite{gherardini2020catheter} as well as tubular structures, such as the whole mouse brain vasculature \cite{todorov2020machine}, the dermal vasculature \cite{gerl2020distance} and neurons \cite{paetzold2019transfer}.

\noindent\textbf{Physics-based augmentations} Another strategy to mitigate the impact of sparse data is image augmentation, where additional training samples are generated based on heuristic transformations of existing data \cite{shorten2019survey}. However, many augmentations used for natural images do not translate to the medical domain as the underlying data acquisition processes are different. This motivates the need for transformations that emulate the physics of medical image acquisition. Such specialized augmentations have been developed for computed tomography \cite{omigbodun2019effects}, magnetic resonance tomography \cite{shaw2020k} and ultrasound imaging \cite{tirindelli2021rethinking}. Even though the physics behind OCTA image acquisition is well understood \cite{spaide2015image}, data augmentation schemes specific to this modality have not been introduced to date.

\section{Methods}
\label{sec:methods}

\subsection{Physiology-based simulation of the retinal vasculature}

\begin{figure}[htbp]
\centering
\includegraphics[width=0.9\textwidth]{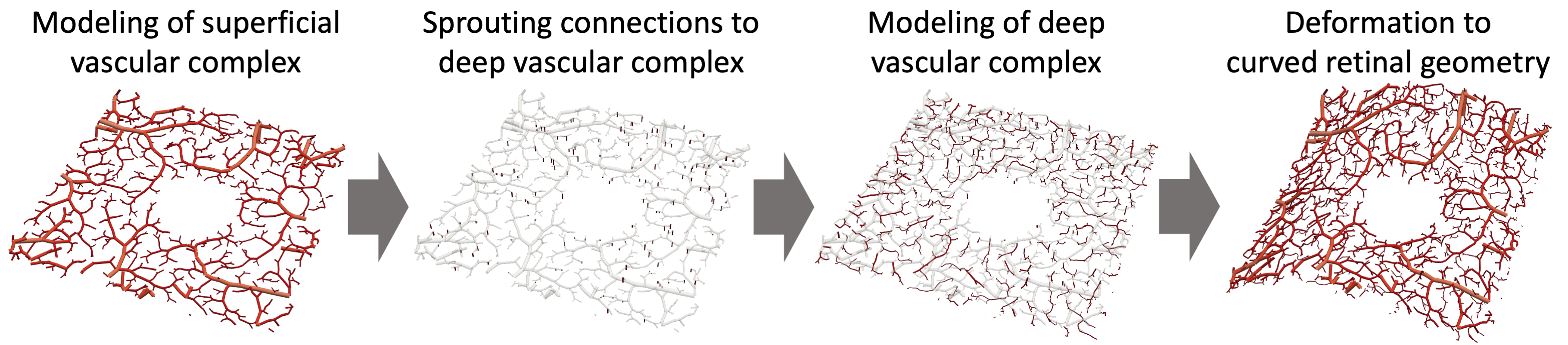}
\caption{Main steps during the modeling of retinal vessel graphs. Additional examples can be found in the supplementary material.}
\label{fig:vascular_model}
\end{figure}

In order to generate diverse and realistic models of the retinal vasculature, we adapt and extend a physiology-based simulation by Schneider \textit{et al.} \cite{schneider2012tissue}. In the following, we briefly describe their method before presenting our key novel contributions that allow us to model the highly unique vasculature of the retina (see figure \ref{fig:vascular_model}).

In Schneider \textit{et al.}'s simulation, blood vessels are represented as a forest of rooted tree graphs. Each graph node encodes a vessel segment modeled as a cylindrical tube with a specific radius, length and position as well as connectivity information to its parent vessel and up to two children. These trees are grown iteratively in a multi-scale fashion. At each simulation step, the concentration maps of oxygen and vascular endothelial growth factor (VEGF) are calculated depending on the spatial configuration of the vessel trees. High VEGF concentration, which drives the growth of new blood vessels, will primarily occur in regions of low perfusion. To counteract the lack of oxygen, new vessels develop from nearby graph nodes via sprouting, elongation or bifurcation. Vessel formation adheres to rules derived from fluid dynamics that enforce morphologically plausible branching statistics.

\noindent\textbf{Sequential simulation of the retinal vascular complexes} The inner retina is supplied with blood by four vascular plexuses, which can be grouped into two complexes: the supervicial vascular complex (SVC) and the deep vascular complex (DVC) \cite{campbell2017detailed,joussen2007retinal}. We sequentially simulate these vascular complexes according to their order in blood stream direction.

The SVC features several large arterioles that branch out as they run towards the center of the retina. Its simulation is initialized with large vessel stumps that are radially distributed along the edges of the volume. The SVC leads to the DVC, which primarily consists of smaller, sprawling vessels. After the SVC's simulation has converged, we calculate the chance of each vessel node to form a vertical sprout following the global branching statistics \cite{schneider2012tissue}. These sprouts are used as initial seeding vessels for the simulation of the DVC. We have heuristically tuned the simulations' hyperparameters so that the generated vasculature models mimic the morphological statistics of the SVC and DVC, respectively. We provide the two sets of hyperparameters and rationale for these settings in the supplementary material.

\noindent\textbf{Modeling of regions with low angiogenesis} The outer retinal layers, which are responsible for the sensing of light, are not directly perfused. Furthermore, all retinal plexuses feature a region devoid of blood vessels at their center, the foveal avascular zone, so that light can pass through to the photoreceptor layers \cite{tick2011foveal}. The lack of angiogenesis in these regions is induced by enforcing a reduced secretion of VEGF using binary masks. The shape of these masks is based on previously reported sizes for the retinal layers and the foveal avascular zone \cite{samara2015correlation}.

\noindent\textbf{Deformation to the retinal layer structure} We model both vascular complexes inside thin rectangular volumes before deforming the vessel graphs to the typical curved shape of the retina as it appears in OCT and OCTA images. To this end, we use the IOWA reference algorithms to extract retinal layer segmentations from anatomical OCT scans contained in the OCTA-500 dataset  \cite{abramoff2010retinal,garvin2009automated,li2005optimal,li2020ipn}. The SVC is sheared to adhere to the shape of the ganglion cell layer. The DVC is deformed to the relief of the inner nuclear layer. Afterwards, we remodel the entire vascular tree one final time in order to enforce physiologically correct branching angles and vessel thinning along blood flow direction.

\noindent\textbf{Conversion to gray-scale images} By running the simulation with varying starting conditions, retinal layer shapes and random number generator seeds, we are able to obtain a diverse set of realistic retinal vessel graphs. The graphs are converted to voxelized gray-scale images with matching ground truth labels. For the following experiments we generate both two-dimensional \textit{en-face} images (i.e. maximum intensity projections onto the axial plane) and three-dimensional image volumes.

\subsection{Image augmentations based on the OCTA acquisition process}

\begin{figure}[htbp]
\centering
\includegraphics[width=0.9\textwidth]{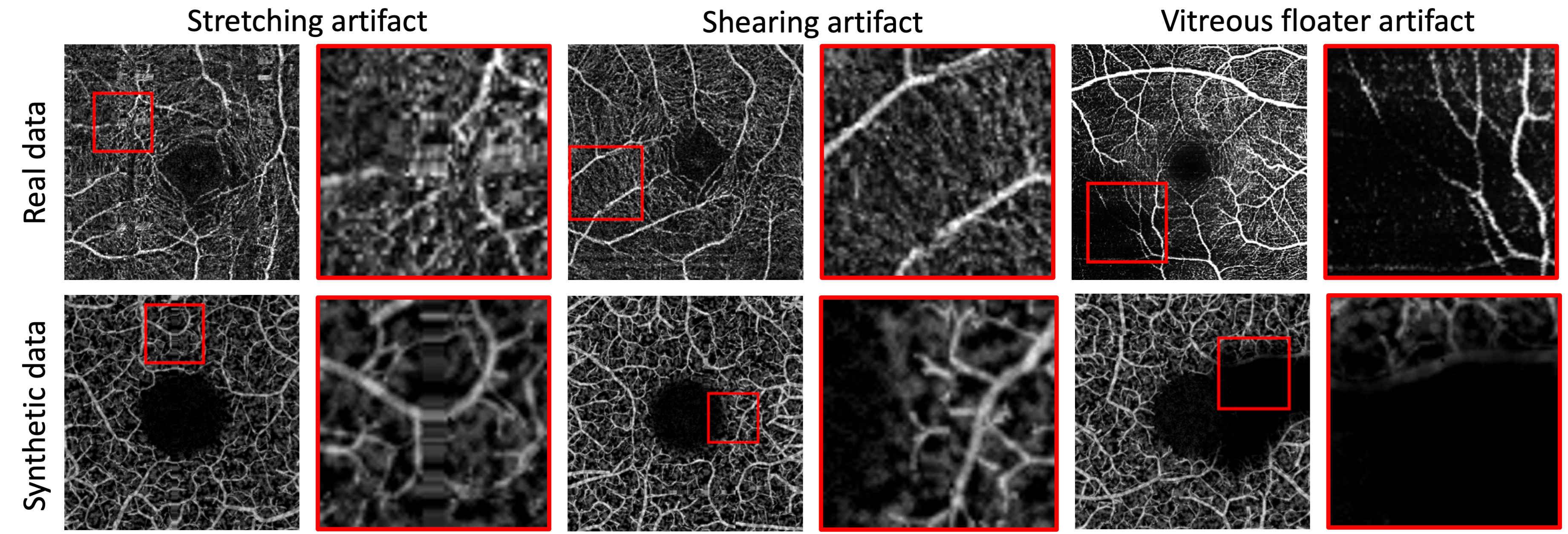}
\caption{\textit{En-face} OCTA images and close-up views of common artifacts in real OCT images from the ROSE dataset \cite{ma2020rose} and representative samples of our synthetically generated dataset. Additional examples can be found in the supplementary material.}
\label{fig:physics_artifacts}
\end{figure}

Next, we apply a series of newly developed image transformations that simulate image features and artifacts introduced during OCTA acquisition (see figure \ref{fig:physics_artifacts}). We briefly describe these physics-based augmentations below, while providing pseudo-code and hyperparameters in the supplementary material.

\noindent\textbf{Flow projection artifacts} OCTA resolves the temporal variation of successive scans of the same volume. Blood cells flowing through the vessels will display as bright signal, while stationary tissue will appear dark. However, as the imaging beam traverses the retina, blood vessels will cast a shadow onto the deeper parts of the retina. This shadow is detected as temporally changing signal and wrongly appears as blood flow in the calculated angiograph \cite{spaide2015image}. We simulate such artifacts by identifying all vessels inside the volume that exceed a certain radius. Subsequently, we increase the image brightness along the projection of these vessels in the imaging beam's direction.

\noindent\textbf{Eye motion artifacts} Volumetric OCTA scans are usually acquired in a rasterized fashion, in which multiple B-scans and small imaging patches are merged together in a postprocessing step. Movement of the subject's eye or head during imaging may result in discontinuities between adjacent B-scans or so-called quilting artifacts at the border of the patches \cite{spaide2015image}. We simulate four different motion artifacts along random horizontal and vertical cuts of the synthetic images: shearing, stretching, buckling and whiteout. The first three are modeled by shifting the intensity values on one side of the cut perpendicular to the imaging direction. The latter one is modeled as uniform noise along the cut and emulates a complete signal decorrelation in a single B-scan.

\noindent\textbf{Vitreous floater artifacts} During acquisition of retinal OCTA images, the light beam and its reflection pass through several transparent components of the eye, including the cornea, pupil, lens and vitreous. Opaque objects inside these structures, in particular vitreous floaters, can cause a loss of spatial coherence of the imaging beam that manifests itself as darkened areas in the image \cite{spaide2015image}. We model the typical thread-like shape of vitreous floaters as a graph of multiple connected line segments with random thickness, length and orientation. We determine the projection of the resulting polygon in imaging direction and reduce the signal in this region.

\noindent\textbf{Background signal caused by capillary vessels} At its smallest scale, the circulatory system branches into numerous capillary vessels. OCTA scanners are typically unable to resolve these small blood vessels, which appear as milky background signal in the image \cite{spaide2018optical}. It is computationally infeasible to simulate the entire capillary vasculature, so that the background of our synthetic images appears unrealistically empty. To account for this limitation, we add binomial noise, which is blurred by a Gaussian filter.

\section{Experiments and results}
\label{sec:experiments}

We evaluate the utility of our synthetic data in several benchmarking experiments using the publicly available ROSE-1 dataset \cite{ma2020rose}. We train supervised vessel segmentation algorithms using either synthetic OCTA image-label pairs ("synthetic") or real OCTA images with expert-derived labels ("real"). For all experiments we assess the algorithms' performance on real OCTA images by calculating accuracy (ACC), area under the receiver operating characteristic curve (AUC), F1-score (DICE) and the topology-aware centerline DICE (clDICE) \cite{shit2021cldice}. We adapt network implementations, data preprocessing and hyperparameter settings from Ma \textit{et al.}\footnote{https://github.com/iMED-Lab/OCTA-Net-OCTA-Vessel-Segmentation-Network}. Diverging from their approach that uses a fixed number of training epochs, we introduce an independent validation split for model selection. As a result the absolute performance and relative rankings of the vary algorithms may slightly differ compared to their reported results. All results are reported as mean and standard deviation across five cross-validation splits.

\begin{table}[htbp]
\caption{Performance of four segmentation networks trained on either synthetic image-label pairs or real images with expert-derived annotations.}
\label{tab:quantitative_results}
\centering
\begin{tabular}{|l|l|l|l|l|l|}
\hline
Model & Training data & ACC & AUC & DICE & clDICE \\ \hline
U-Net \cite{ronneberger2015u} & Real & 0.89$\pm$0.01 & 0.91$\pm$0.01 & 0.69$\pm$0.02 & 0.60$\pm$0.02 \\
 & Synthetic & 0.80$\pm$0.02 & 0.81$\pm$0.01 & 0.54$\pm$0.02 & 0.47$\pm$0.02 \\ \hline
ResU-Net \cite{zhang2018road} & Real & 0.88$\pm$0.01 & 0.90$\pm$0.01 & 0.68$\pm$0.02 & 0.59$\pm$0.02 \\
 & Synthetic & 0.84$\pm$0.01 & 0.82$\pm$0.02 & 0.56$\pm$0.03 & 0.48$\pm$0.04 \\ \hline
CS-Net \cite{mou2019cs} & Real & 0.89$\pm$0.01 & 0.90$\pm$0.01 & 0.69$\pm$0.01 & 0.59$\pm$0.02 \\
 & Synthetic & 0.77$\pm$0.04 & 0.81$\pm$0.02 & 0.55$\pm$0.02 & 0.48$\pm$0.02 \\ \hline
CE-Net \cite{gu2019net} & Real & 0.88$\pm$0.01 & 0.88$\pm$0.01 & 0.67$\pm$0.02 & 0.58$\pm$0.02 \\
 & Synthetic & 0.85$\pm$0.02 & 0.84$\pm$0.02 & 0.59$\pm$0.02 & 0.51$\pm$0.04 \\ \hline
\end{tabular}
\end{table}

\begin{figure}[htbp]
\centering
\includegraphics[width=0.85\textwidth]{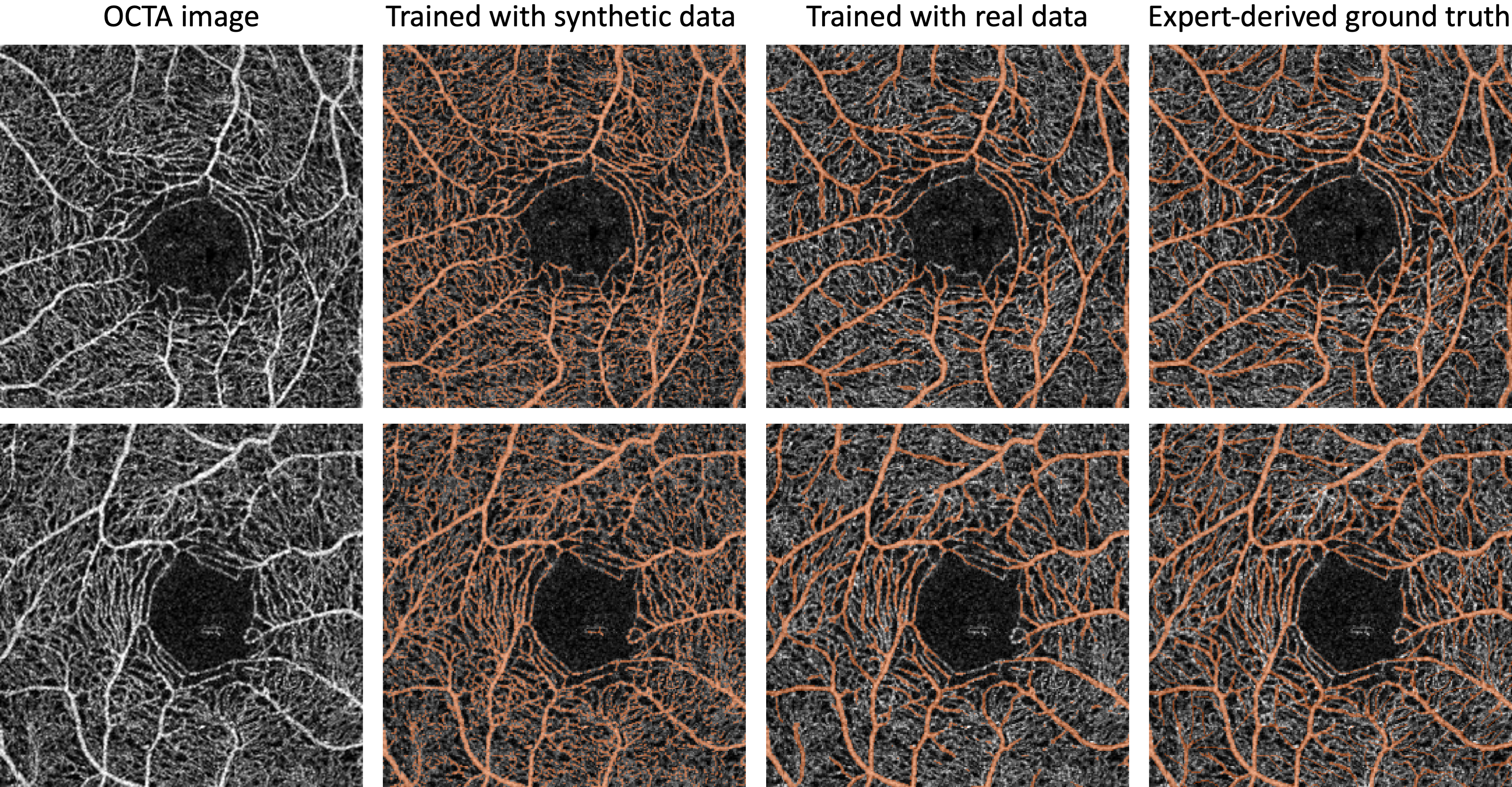}
\caption{Two representative test samples with segmentation maps overlaid in orange. Algorithms trained on synthetic data accurately segment capillary vessels.}
\label{fig:qualitative_results}
\end{figure}

We find that the use of synthetic data facilitates vessel segmentation in OCTA images across all four considered neural network architectures. However, the quantitative scores are consistently lower than those of networks that were trained on the same domain (see table \ref{tab:quantitative_results}). Qualitatively, we find our segmentations to be superior (see figure \ref{fig:qualitative_results}). We suspect that this discrepancy is caused by a systematic difference in the datasets' annotations. The expert labels in the ROSE-1 dataset do not encompass the smallest capillaries, whereas these vessels are labeled in our synthetic dataset. Consequently, the networks trained on synthetic data label these important small vessels, whereas the baselines do not.

An advantage of our synthetic data pipeline is that it can easily generate large amounts of training data. In additional experiments, we show that network performance increases with training dataset size (see table \ref{tab:scaling_results}). We also observe that top-end network performance improves when pretraining networks on synthetic data before finetuning them on a small number of real samples. In further ablation experiments, we quantified the benefit of both newly introduced components. Using vessel trees without the typical structure of the SVC and DVC results in a decreased segmentation accuracy (see table \ref{tab:ablation_results}). Similarly, we find that leaving out the physics-based image transformations and training only with generic augmentations reduces performance.

\begin{table}[htbp]
\caption{U-Net performance across different dataset types and training strategies.}
\label{tab:scaling_results}
\centering
\begin{tabular}{|l|l|l|l|l|l|}
\hline
Training data & Dataset size & ACC & AUC & DICE & clDICE \\ \hline
Synthetic & 32 & 0.80$\pm$0.02 & 0.81$\pm$0.01 & 0.54$\pm$0.02 & 0.47$\pm$0.02 \\
Synthetic & 320 & 0.81$\pm$0.02 & 0.83$\pm$0.01 & 0.56$\pm$0.01 & 0.48$\pm$0.02 \\ 
Synthetic & 3200 & 0.81$\pm$0.01 & 0.84$\pm$0.01 & 0.57$\pm$0.01 & 0.49$\pm$0.02 \\ \hline
Synthetic $+$ finetuning & 352 & 0.91$\pm$0.01 & 0.91$\pm$0.01 & 0.71$\pm$0.01 & 0.62$\pm$0.02 \\ \hline
Real & 32 & 0.89$\pm$0.01 & 0.91$\pm$0.01 & 0.69$\pm$0.02 & 0.60$\pm$0.02 \\ \hline
\end{tabular}
\end{table}

\begin{table}[htbp]
\caption{U-Net performance after ablating the physiology-based simulation of the retinal vasculature or the physics-based augmentations of our data synthesis method.}
\label{tab:ablation_results}
\begin{tabular}{|l|l|l|l|l|}
\hline
Configuration & ACC & AUC & DICE & clDICE \\ \hline
Proposed method & 0.81$\pm$0.02 & 0.83$\pm$0.01 & 0.56$\pm$0.01 & 0.48$\pm$0.02 \\
$-$vessel simulation & 0.79$\pm$0.01 & 0.80$\pm$0.01 & 0.53$\pm$0.01 & 0.45$\pm$0.01 \\ 
$-$augmentations & 0.79$\pm$0.01 & 0.70$\pm$0.02 & 0.52$\pm$0.01 & 0.47$\pm$0.01 \\ \hline
\end{tabular}
\centering
\end{table}

\begin{figure}[htbp]
\centering
\includegraphics[width=0.725\textwidth]{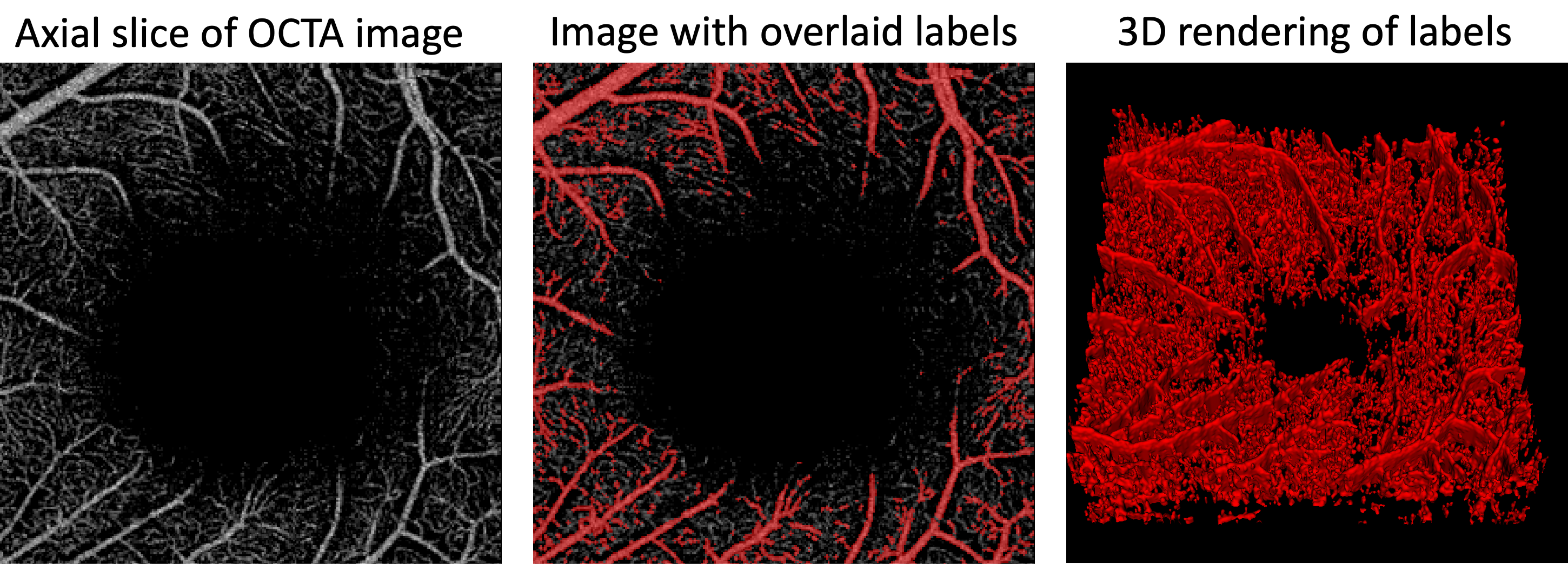}
\caption{Three-dimensional vessel segmentation of a volumetric OCTA image by a U-Net trained exclusively on our synthetic data.}
\label{fig:3d_results}
\end{figure}

Finally, we train a 3D U-Net for vessel segmentation in volumes of the OCTA-500 dataset using only synthetic training data \cite{li2020image,li2020ipn}. As of now, we are unable to benchmark on this task as there are OCTA datasets with volumetric labels are not available. Instead, we present initial qualitative results (see figure \ref{fig:3d_results}). These highlight the potential of our method to facilitate three-dimensional analysis of OCTA images; a task with enormous diagnostic potential which can only be rarely conducted due to a lack of suitable tools \cite{yu2021cross}.

\section{Discussion  and conclusion}
\label{sec:discussion}

In this work we have presented a method to generate highly realistic, synthetic OCTA images. Our method models the retinal vasculature with excellent anatomical detail using a physiology-based angiogenesis model. Moreover, we simulate image features and artifacts resulting from the OCTA acquisition process using a set of physics-based image augmentations. While our synthetic OCTA images are still distinguishable from real images, they are accompanied by a fully characterized vascular graph. This intrinsically matched ground truth is highly advantageous when using the dataset for downstream tasks.

We demonstrate the promise of our approach by successfully training several segmentation algorithms without manual ground-truth annotations. In the future, our method can be tailored to emulate the characteristics of diverse clinical datasets. The vasculature simulation can be adapted to model the retinal morphology of specific populations, such as older subjects or patients with ocular, cardiac or neurological pathologies \cite{joussen2007retinal}. The image augmentations can be tuned to the hard- and software build into different OCTA scanners \cite{spaide2018optical}. We believe that this versatility will enable application of our method beyond vessel segmentation and ultimately advance the quantitative analysis of OCTA in clinical practice.

\section*{Acknowledgments}
Johannes C. Paetzold is supported by the DCoMEX project, financed by the Federal Ministry of Education and Research of Germany. Benjamin Knier is funded by the Else Kröner-Fresenius-Stiftung, the Gemeinnützige Hertie Foundation and received a research award from Novartis.

\bibliographystyle{splncs04}
\bibliography{bibliography}
\clearpage

\appendix
\chapter*{Supplementary material}

\setlength{\intextsep}{1.25\baselineskip plus 0.0\baselineskip minus 0.0\baselineskip}

\begin{figure}[htbp]
\centering
\includegraphics[width=0.9\textwidth]{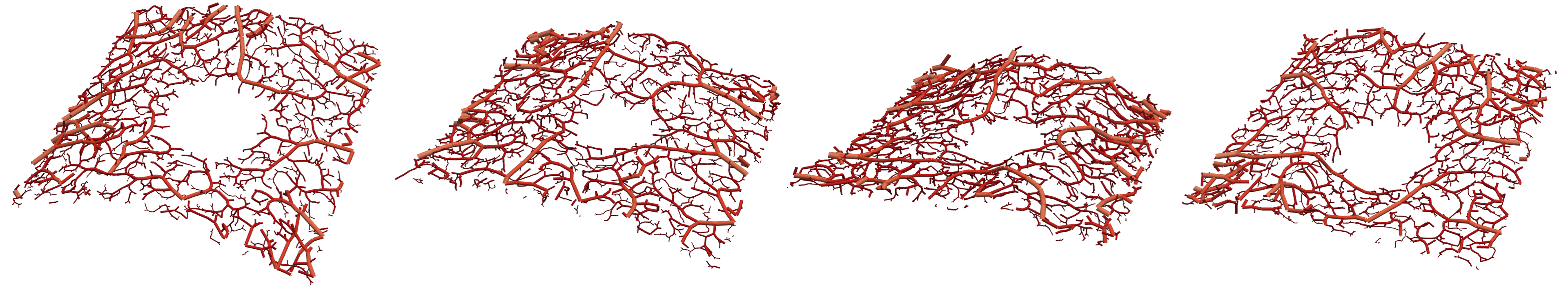}
\caption{Random examples of synthetically generated retinal vessel graphs.}
\end{figure}

\begin{table}[htbp]
\scriptsize
\caption{Tuned hyperparameters for the simulation of retinal vessel graphs. The changes compared to the implementation by Schneider \textit{et al.} (original) facilitate the synthesis of the superior vascular complex (SVC) and deep vascular complex (DVC).}
\begin{tabular}{|l|l|l|l|l|}
\hline
Parameter & original & SVC & DVC & Rationale behind change \\ \hline
$\Omega_x$ & \SI{936}{\micro\meter} & \SI{3200}{\micro\meter} & \SI{3200}{\micro\meter} & Flat simulation volume for modeling of retinal vasculature \\
$\Omega_y$ & \SI{864}{\micro\meter} & \SI{3200}{\micro\meter} & \SI{3200}{\micro\meter} & Flat simulation volume for modeling of retinal vasculature \\
$\Omega_z$ & \SI{1600}{\micro\meter} & \SI{64}{\micro\meter} & \SI{64}{\micro\meter} & Flat simulation volume for modeling of retinal vasculature \\
$\theta_c$ & 1.050 & 1.025 & 1.025 & Encourage more sprouting of vessels \\
$\theta_p$ & 0.95 & 0.90 & 0.90 & Faster model convergence without noticeable loss of detail \\
$r\textsubscript{initial}$ & \SI{15.0}{\micro\meter} & \SI{22.5}{\micro\meter} & $-$ & Larger initial radius of the retinal arterioles \\
$r\textsubscript{min}$ & \SI{3.0}{\micro\meter} & \SI{2.25}{\micro\meter} & \SI{1.5}{\micro\meter} & Simulation of smallest retinal capillaries \\
$r\textsubscript{degen}$ & \SI{5.00}{\micro\meter} & \SI{3.75}{\micro\meter} & \SI{2.50}{\micro\meter} & Simulation of smallest retinal capillaries \\
$r\textsubscript{prune}$ & \SI{2.00}{\micro\meter} & \SI{1.5}{\micro\meter} & \SI{1.00}{\micro\meter} & Simulation of smallest retinal capillaries \\
$m_b$ & 14 & 16 & 12 & Longer/ shorter vessel segments before bifurcation \\ 
$\lambda_g$ & 1.00 & 1.00 & 0.25 & Contorted, web-like structure of DVC \\
$\gamma$ & 3.0 & 3.0 & 2.5 & Larger bifurcation angles in DVC \\
\hline
\end{tabular}
\centering
\end{table}

\begin{figure}[htbp]
\centering
\includegraphics[width=0.9\textwidth]{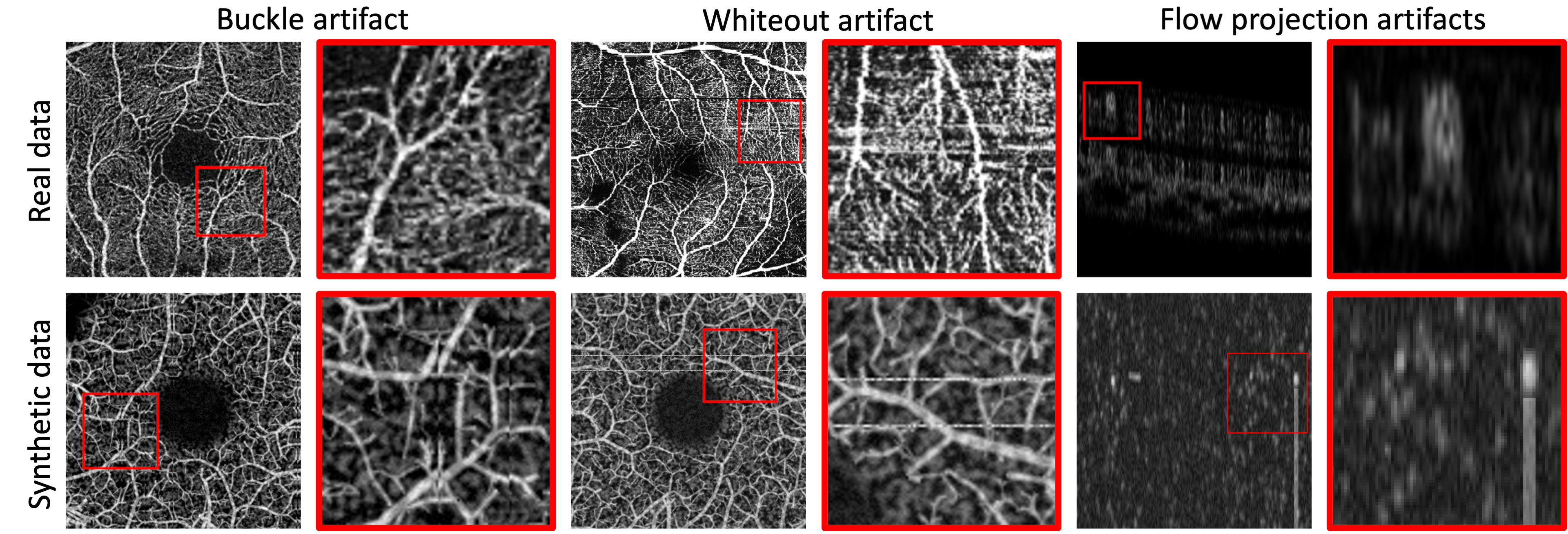}
\caption{Three additional examples of artifacts in real and synthetic OCTA images.}
\end{figure}

\clearpage
\noindent \textbf{Pseudo-code for the physics-based image augmentations.} The default hyperparameters used in this study are included in the algorithms' requirements. Units of size are provided in pixels/ voxels unless explicitly specified.

\begin{figure}[htbp]
\centering
\includegraphics[width=0.925\textwidth]{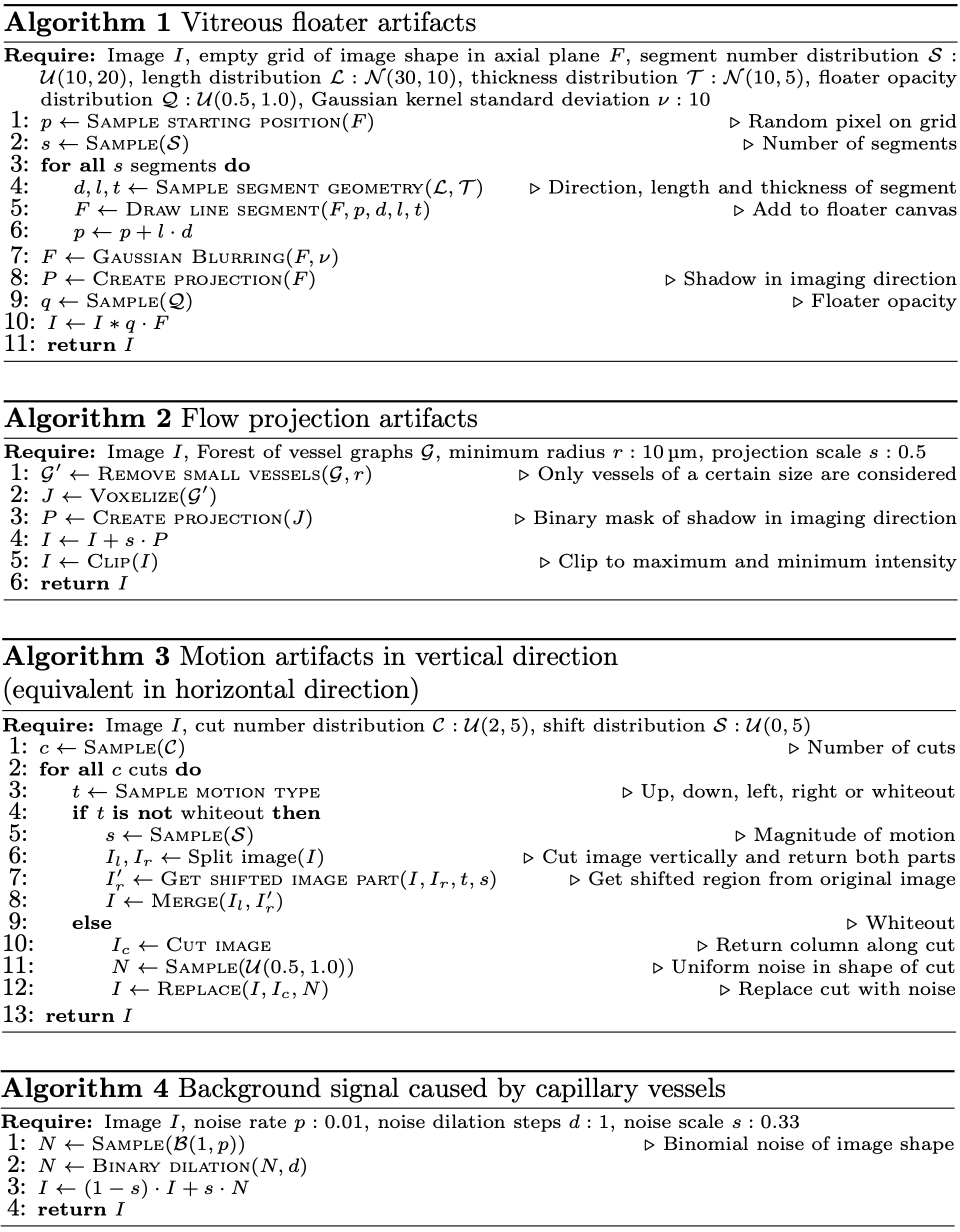}
\end{figure}

\end{document}